# Exploring Collaborative Debating Models for the future of science

Zeki C. Seskir

Institute for Technology Assessment and Systems Analysis (ITAS),

Karlsruhe Institute of Technology (KIT), Karlsruhe, Germany

## Abstract

AI is expanding what researchers can produce. A substantial increase in scientific output would also change how that work is organised and distributed, as earlier shifts in the scale of science have done. This paper develops a thought experiment around Collaborative Debating Models (CDMs), topic-focused environments where researchers work with AI and one another to pursue questions and maintain an evolving account of their field. Individual contributions would remain attributable as ongoing debate and further analysis change the shared account. The paper considers how journals could evolve into these gathering places and how different forms of ownership would shape scientific agendas and the support for continuing work. CDMs could connect participants to remote laboratories and other research services to conduct new studies, with shared repositories making the resulting data available for further inquiry. The proposal links expanded scientific production to collective interpretation through continuing assessment and revision, while leaving room for disciplinary differences in how contributions are reviewed and recognised.

## 1 A change in the scale of science

On 6 October 2026, OpenAI released more than 700 mathematical manuscripts as preprints, produced by an internal AI model (OpenAI 2026a, 2026b). The release sent shockwaves through the mathematical community, prompting fierce debate about the future of the discipline and the place of researchers within it (Hart 2026). On 26 October, the company announced further results extending into theoretical physics and invited academic groups to turn its outputs into scientific understanding (OpenAI 2026c).

Similar seismic shifts are expected to reach other academic disciplines over the coming months and years. In computational social science, a recent review already describes large language models (LLMs) assisting analysis and hypothesis development, while AI agents themselves increasingly become subjects of research (Xu et al. 2026). The pace of change will vary with the work involved, particularly where progress depends on collecting new evidence, but the institutional question is already with us: how will science organise a major increase in what researchers can produce?

Science has changed scale before. In *Little Science, Big Science*, Price (1963: vii–viii) described the growth of scientific activity and the arrangements for administering it that had developed as needs arose. His analysis helps us see why a change in productive capacity reaches beyond the individual researcher: a larger scientific enterprise needs ways to organise its work and circulate the results, and the institutions built for that purpose acquire influence over what science becomes.

The prospect of further growth also comes amid concern that scientific progress has become harder to sustain. Bloom et al. (2020: 1104–1105) provide an influential account of this problem, finding that rising research effort accompanied declining research productivity across the industries and firms they examined. The concern is about how much effort progress requires, even as scientific output continues to expand. Scientific publishing has meanwhile become a major focus of criticism, particularly over commercial control and the incentives attached to journal prestige. Brembs et al. (2023: 2–6) connect these problems to the organisation of scholarly communication and argue for replacing journals with infrastructure governed by scholars. AI enters this debate at a moment when both the production of knowledge and the institutions through which it circulates are already being questioned.

Collaborative platforms offer useful precedents for thinking beyond the individual paper. The Polymath Project brought mathematicians together through open discussions, with a wiki drawing their contributions into a developing account of the problem (Gowers and Nielsen 2009: 879). The Open Research Knowledge Graph (ORKG) supports structured comparisons of research contributions, maintained through expert curation and connected to research workflows (Stocker et al. 2023: 20–21). These approaches have become part of a wider landscape of platforms supporting scientific work. Da Silva Neto and Chiarini (2023: 3–4, 18) describe this development as the platformization of science, showing how platforms organise interactions across research activities and become embedded in scientific practice. Their importance lies partly in bringing people together around work that continues between publications.

Ethan Mollick's *Co-Intelligence* (2024) provides a useful starting point: AI can become a partner in thought, with people developing ideas through the exchange and judging what to make of its outputs. One of his principles is to "assume this is the worst AI you will ever use" (Mollick 2024, ch. 3). We should expect better, probably much better, systems in the coming years. Yet Mollick also argues that even if development stopped, the capabilities already achieved would still have far-reaching consequences (2024: xvii). Even without further breakthroughs, researchers would have these capabilities to build on, and institutions would have to decide how their use should be organised.

Lipshitz's (2026: 31–34) account of using a personal AI subscription gives this possibility a concrete form. In his knot-theory preprint, he describes returning to questions he had set aside, using a $20-per-month Claude Pro subscription. Models carried out much of the research, while he guided their work and spent 50–100 hours checking and rewriting the proofs. His requests for alternative arguments also improved the mathematics. Co-intelligence made room for work he had not found time to pursue, with his judgement shaping what became a paper.

Let us therefore entertain a future in which AI enables researchers to pursue substantially more science. Researchers remain central: they choose questions and bring the experience needed to judge what an answer means, while AI expands the work they can undertake. The resulting growth would create a major distribution problem. Making another million papers available would leave open how researchers could find what mattered and understand how the contributions fitted together. This is the starting point for the thought experiment: Collaborative

Debating Models, or CDMs, through which scientific communities could organise that expanded production and make sense of its results together.

## 2 Production distribution and scientific authority

Scientific journals have long connected the circulation of findings to the organisation of a research community. The first issue of *Philosophical Transactions* appeared in 1665, providing a continuing vehicle for communicating scientific observations and discoveries (Fyfe et al. 2014: 4, 7–8). As the number of journals and papers grew, specialised venues helped researchers identify relevant work and locate themselves within a field. Publication also established a visible record of contribution on which claims to priority and reputation could rest. The journal came to organise both the movement of knowledge and the recognition of those producing it.

Researchers have used publication records to examine science for more than a century. As early as 1917, Cole and Eales (1917: 578) counted publications to trace the development of comparative anatomy. Such work anticipated scientometrics, the quantitative study of science through records such as publications and citations. Price's (1963) analysis of scientific growth illustrates how the literature can become evidence about science itself. Once publications could be counted and their citation relationships traced, these records also offered institutions convenient ways to compare researchers. Indicators such as the h-index turned patterns of publication and citation into portable academic credentials, making a numerical account of a career available far beyond the specialists able to judge its work.

That convenience carries power. Hicks et al. (2015: 429–431), in the Leiden Manifesto, explain how metrics enter decisions about careers and research funding, while their meaning depends on disciplinary practices and the coverage of the databases used. A score can follow a researcher into a hiring decision even when the people reading it know little about the conditions that produced it. Researchers, in turn, have reasons to pursue work likely to improve the score. An infrastructure initially used to describe scientific activity can thereby help direct it.

Citation patterns also reflect the distribution of attention. Merton's (1968: 57–59) account of the Matthew effect showed how established scientists could receive disproportionate recognition, with reputation helping their work become visible and supporting access to further resources. Growth does not remove this problem; a larger literature can make familiar names and venues more attractive as guides. Publishing more work therefore leaves the question of who will encounter it, and whose contributions will be treated as worth following, firmly in place.

AI could produce a similar shift much faster. If researchers increasingly encounter a field through an AI-generated explanation, choices about the sources it includes and the disagreements it highlights will affect which work receives attention. A system that helps generate research may also become the system through which others discover and interpret it. This makes distribution inseparable from production. Then the question becomes who will organise those encounters and what authority researchers will have over them. CDMs let us explore how communities could retain influence over these encounters as research production expands.

## 3 Collaborative Debating Models

A Collaborative Debating Model is a proposed way for a research community to organise its work through a shared AI-supported environment. Think of it as a topical journal meeting Zotero, arXiv and Wikipedia. It could bring together several underlying AI systems, with researchers using them to investigate questions and develop a continuing account of the work. The term “debating” describes the community’s engagement with competing explanations, including exchanges assisted by agents. Here, an agent is an AI model given a task and access to relevant materials or computational tools. The CDM provides a gathering ground where participation can continue through the inquiry itself and after a result has appeared.

For a moment, suspend disbelief about how quickly such an environment could become practical. Imagine opening the CDM for your field and finding an account of a question you want to pursue, together with the contributions supporting it and the reasons a disagreement remains unresolved. You can follow the argument back to the original work, then invite other researchers to examine a particular difficulty. AI assistance makes it easier to reconstruct methods or run a calculation while the discussion develops. What emerges can change the account that the next researcher encounters, rather than disappearing into correspondence once the immediate collaboration ends.

A few familiar services may help explain the arrangement. A shared Zotero library lets a group collect references and organise material for its work (Zotero 2024). arXiv provides identifiable contributions whose earlier versions remain available as authors revise them (arXiv n.d.), while Wikipedia shows how people can maintain an account through continuing collective revision (Forte et al. 2009). A CDM would connect these activities: original contributions would remain attributable, while the collective account explained how they related and changed as further work appeared. Assessing original scientific claims would require disciplinary procedures beyond Wikipedia's synthesis of existing sources. Researchers could agree that a dispute had been represented fairly while continuing to disagree about its resolution.

Parts of this arrangement already exist. ORKG's expert observatories curate comparisons around research problems, allowing researchers to examine how contributions have been represented (Stocker et al. 2023: 20–21). Paper2Agent makes papers and their associated resources available through interactive agents, including cooperation between agents representing different papers (Miao et al. 2026, "AI co-scientists through agent collaboration"). These are useful precursors because they let published work become material for further inquiry. The thought experiment connects that activity to a community that takes responsibility for the account developing around it, extending the relationship between discussion and synthesis already visible in Polymath.

Let's think of a scenario. Two studies appear to disagree about the effects of an intervention. Within a CDM, a researcher asks an agent to reconstruct the studies' assumptions, while a colleague familiar with the methods examines how the outcomes were measured. They discover that the comparison has treated different populations as equivalent. A further analysis, undertaken with the relevant data and permissions, shows that the apparent contradiction turns on that difference. Researchers from the two groups can then debate whether the collective account should narrow the claim or identify a question requiring a new study. The thought experiment extends Mollick's (2024) co-intelligence to a research community: researchers redirect the analysis as their understanding develops, and the agent's results become material for colleagues to interpret and question together.

The resulting explanation would enter the account with its supporting analysis and the reasons for the revision. The original studies would retain their attributed findings, while later readers could see why their relationship had been reinterpreted. Editors or an agreed specialist group would take responsibility for the change, with a route for contributors to challenge it. Longino's (1996: 40–41) account of critical interaction is useful here: scientific criticism needs both a forum and the possibility of affecting accepted understanding. A CDM would make that connection part of its everyday organisation. Its value would depend on whether a well-founded objection could change what the community presented as known.

MetaResearch Open Review (MetaROR) offers a practical precedent. It organises open peer review of metaresearch articles deposited in repositories, publishing editorial assessments alongside the reviews and retaining author responses with links to revised versions. These assessments can accompany later journal submissions (MetaROR n.d., “Peer review process”). A CDM could build on this separation of assessment from journal publication, connecting discussion of individual contributions to a continuing account of the field.

Wikipedia’s editorial work combines shared rules with topic-specific groups (Forte et al. 2009). Community Notes on X uses agreement among contributors with different past rating patterns to identify helpful notes (X n.d.). These arrangements offer ways to organise scrutiny across disagreement, with a CDM adapting that organisation to the expertise and assessment practices of its research community.

As new contributions arrived, agents could suggest which existing questions they affected and assemble comparisons for researchers to examine. The resulting assessments would update the shared account, giving readers a way to see what had changed without reconstructing the whole literature themselves. A finding could reach colleagues through the question they were following as well as through its own publication. Maintaining that account would concentrate some of the interpretive work in a common resource, where later contributions could correct or extend it.

A CDM could also make more kinds of contribution visible. An explanation that reconciles two findings could receive attribution even when it does not warrant a full article, and maintaining a useful comparison could become recognised scientific work. Periodic versions of the account could be published and cited, giving readers a stable reference while discussion continued. Researchers entering from another domain could encounter both an explanation and the people responsible for it. In this sense, a CDM would distribute opportunities to contribute alongside the results of previous work.

The proposal leaves room for different disciplinary cultures. A CDM centred on mathematical arguments might develop through exchanges that participants could inspect directly, while experimental communities would remain dependent on laboratories and access to evidence held elsewhere. Topic-focused communities could maintain distinct environments and cooperate where their questions overlap. Such connections would let specialised communities develop their own questions while drawing on one another's work.

## 4 Who would run a CDM

Da Silva Neto and Chiarini (2023: 20–21) show how scientific platforms take shape through different institutional origins. For CDMs, ownership would put control over scientific agendas and the resources for continuing work in different hands. The four routes below distinguish who owns and governs the institution; public funding or commercial services could support more than one of them.

### Publishing houses

One route would run through scientific publishing houses. A topic-focused journal could gradually become the centre of a CDM, drawing on its editors and existing community of authors and readers. Discussion around published work could develop into maintained comparisons, with agents helping participants investigate the questions arising from them. Over time, the journal's periodic publications might become selected versions of a continuing scientific conversation. Each CDM could develop its own culture and standards, much as journals and their communities already differ.

The money supporting scientific publishing offers a possible basis for this transition. Brembs et al. (2023: 9) argue that existing scholarly communication expenditure could finance a different infrastructure. Applied to CDMs, this suggests redirecting part of subscription and publication spending towards their development and continuing operation. Institutions would pay for the work of maintaining a scientific environment, including the expertise needed to assess contributions, alongside computing costs. The scale of each CDM would depend on the work its community wanted to sustain and the resources committed to it.

Publisher involvement could make a transition easier where editorial communities and reliable services already exist, but the terms would decide what researchers gained. Participation in debate would not by itself give contributors control over editorial rules or the accumulated account. A publisher could retain both while offering a useful service. A different agreement could give the community authority over scientific decisions and rights to take its work elsewhere. The consequential question would be whether researchers could continue the collective project if their relationship with the provider ended.

### Scholarly associations and academic institutions

A second route would place ownership with scholarly associations or academic institutions. Associations such as the American Physical Society (APS), the Deutsche Physikalische Gesellschaft (DPG), the Institute of Electrical and Electronics Engineers (IEEE), or the European Association for the Study of Science and Technology (EASST) could provide a home for CDMs, supported by membership income or funding agreements. University faculties and research institutes could also build on their topical journals. These arrangements could draw on diamond open access, where neither readers nor authors pay publication charges. The Action Plan for Diamond Open Access connects this approach to sustained institutional support and explicitly addresses ownership and governance (Ancion et al. 2022: 3, 5–6).

*TATuP* offers a concrete example of an institute-led journal serving a specialised community. The Institute for Technology Assessment and Systems Analysis (ITAS) at Karlsruhe Institute of Technology owns and manages it, while oekom publishes it on the institute's behalf (*TATuP* n.d.). A CDM could grow around such a journal's ongoing debates, with the institute providing continuity and the community shaping its scientific direction. The arrangements for editorial appointments would determine how much authority contributors had within that institutional home.

### Funding agencies

A third route would place a CDM under the ownership of a public funding agency, with its operation delegated to a consortium where appropriate. The argument starts from taking science as a public good. Nelson's (1959: 302, 305–306) rationale for public support of basic research concerns benefits that private sponsors cannot fully capture. Applied to CDMs, this supports public investment in a resource whose value grows as others use and develop it. An agency could sustain access and maintenance beyond individual grants, treating the accumulated account as part of the infrastructure available to the wider research community.

Public ownership would make the relationship between agency priorities and scientific independence consequential. Researchers could choose editors through community procedures, while the agency remained accountable for the resources committed. A funder tying support closely to its current programmes would shape the agenda differently from one sustaining a community whose questions changed over time. Long-term support would therefore need arrangements through which researchers could pursue questions beyond the agency's immediate priorities.

Expected project outputs could evolve with these arrangements. A future Horizon Europe or US National Science Foundation funding call could recognise the creation of a CDM, or a substantial contribution to an existing one, among its expected scientific outputs. A project might deliver an assessed comparison that became part of a CDM, with support for keeping it current after the grant ended. Funding would then recognise the collective resource produced through the project alongside its individual publications.

**AI companies**

A fourth route would place CDMs directly under AI companies. The mathematics case shows why this possibility deserves attention: a company can already select research problems and release a substantial body of work around which scientists then organise their activity. In its October 2026 response, the Advisory Group on Mathematics and Artificial Intelligence warned that companies designating researchers to interpret AI outputs could restrict the community's agency and encourage alignment with company priorities (2026, "Denied agency" and "Biased selection"). The concern reaches beyond whether the outputs are useful to how a field comes to spend its time.

OpenAI's 26 October announcement gives external working groups independence over their conclusions and recognises their work of explanation. The groups helped determine release arrangements, while a separate contribution to the Simons Foundation supports independently administered work on AI-generated results (OpenAI 2026c). Such arrangements leave room for company support under academic direction. OpenAI also acknowledges that choosing the initial conveners shapes who participates first and who has early access. For a CDM, the question would be how those initial arrangements developed into continuing control over the research environment.

Direct company control is the least attractive of these routes. If the same organisation controls access to the strongest models and the environment through which a community works, its commercial priorities can become conditions of scientific participation. Research that demonstrates the provider's capabilities may receive privileged support, while other questions become harder to pursue. The community's future could become tied to a product strategy it has little power to influence. An institutional home intended to expand researchers' agency would then become a means of directing that agency towards the provider's interests.

Ownership and the purchase of AI services can be organised separately. Brembs et al.'s (2023: 4–6) distinction between scholarly governance and replaceable service provision offers a useful principle: the organisation supplying a model need not own the scientific institution built around it. Agreements could secure usable exports and rights to continue the work with another provider, supported by people able to maintain the resource through the transition. Publishers and public institutions also have interests that can constrain research. The preferable direction is to keep scientific judgement and the continuity of accumulated work under arrangements that researchers can meaningfully influence, even where commercial providers supply substantial parts of the system.

## 5 Familiar problems in a different setting

A CDM would inherit the scientific community that built it, including its inequalities. Established researchers could dominate an account through their reputation, while less familiar voices struggled to have an objection heard. Merton's (1968) analysis remains relevant because the platform would offer new places for cumulative advantage to operate. Citation counts could become measures of contributions to an account, and researchers might learn to optimise those measures as readily as existing publication indicators. Changing the medium would leave the underlying competition for recognition in place.

AI could reinforce these patterns through the sources and approaches it repeatedly foregrounded. Messeri and Crockett (2024: 53–54) describe the risk of scientific monocultures and illusions of understanding when researchers rely on AI. Within a CDM, an apparently comprehensive explanation could discourage people from examining what had been omitted, especially if several agents repeated the same interpretation. Contributors would need ways to question the comparison categories and the selection of evidence, as well as the accuracy of individual answers. Keeping disagreement visible would be part of maintaining a useful scientific account.

There would also be more work to do. Producing candidate explanations quickly can increase the burden on those deciding which deserve attention, while a community with limited editorial capacity could struggle to keep its account current. Access to powerful models would remain unequal, as would access to the data and equipment needed to test their suggestions. A shared environment could make those dependencies easier to see and coordinate, but resources would still have to be allocated. The enabling premise of the thought experiment depends on funding this continuing work, rather than treating researchers' attention as an unlimited free input.

## 6 Extending scientific participation via CDMs

The same infrastructure could help a community obtain the evidence needed to move an inquiry forward. In the earlier example, identifying a difference between study populations might lead researchers to ask for observations from a group that had been overlooked. Connecting the CDM to existing research services would give that question a route into new work, with opportunities for people beyond the initial collaboration to participate.

For research involving people, companies such as Prolific connect researchers with paid participants, supporting recruitment through prescreening and the use of researchers' own study software (Prolific 2026). A community could develop a study within the CDM and use such a service to recruit participants, with researchers responsible for the design and conditions of participation. Its findings would return to the collective account alongside an explanation of why that study was needed. Data collection would become part of a continuing inquiry that others could understand and extend.

Remote laboratories extend this possibility to physical experiments. Arias and Taylor (2024) describe their use of Emerald Cloud Lab for research on DNA nanostructures, including the practical work of conducting and troubleshooting experiments remotely. Access to such a facility could let a CDM community execute a jointly developed protocol without every contributing group owning the necessary equipment. Shared funding could support access, while experience gained in running the protocol became available to other participants. The community would have to decide which experiments it could afford and how to allocate instrument time, working with the techniques and support the facility offered.

AI agents could also help decide which experiment to run next. Singh (2026, "Level 1 Results") reports giving agents control over a superconducting device's operating settings in a quantum-sensing experiment. Within a fixed measurement budget, they identified promising settings that his coarse manual scan missed; follow-up measurements confirmed the features. Connected to a CDM, such a laboratory could turn a community's question into successive measurements, with researchers interpreting the results and deciding how they changed the shared account.

Repositories such as Zenodo could preserve the data and software generated through this work, with descriptive metadata and separately citable versions (Zenodo n.d.). A CDM could link an explanation to the particular dataset and code used, giving later contributors a starting point for extending the analysis. The FAIR principles would guide how these materials were prepared to be findable, accessible, interoperable and reusable (Wilkinson et al. 2016: 4). Rich metadata would explain their provenance and conditions of reuse, while shared standards would help researchers and agents locate relevant material. Access conditions would remain explicit, including authorisation where needed. Connecting the discussion to such repositories would let evidence travel beyond the original project and support inquiries its contributors had not anticipated.

Participation could also reach beyond established research groups. In Galaxy Zoo, volunteers classified galaxy images, and Hanny van Arkel's identification of an unusual object initiated further astronomical investigation (Lintott et al. 2009). Her contribution shows how someone undertaking a defined task can bring a new question to scientific attention. A CDM could give such observations a continuing home, where contributors followed the investigation and helped interpret what emerged. AI assistance could help newcomers understand the discussion well enough to extend it, alongside engagement with more experienced participants.

The democratic possibility lies in giving more people a part in choosing questions and interpreting results, alongside opportunities to contribute data or carry out a task. A CDM could let participants propose inquiries and help decide which deserved collective resources, with support for those who lacked the time or expertise to take part. Its rules would determine how these contributions entered scientific decisions. Widening participation in this way could change whose concerns guided the inquiry and whose observations led it in a new direction.

## 7 A gradual transition and open questions

Such a transition would not happen overnight. CDMs could begin alongside existing publication practices, with a journal attaching a maintained account to an ongoing debate while authors continued to publish individual papers. There is ample room for experimentation in different settings, from a small institute's journal to a community supported by a national funder. Participants could develop the arrangement as their collaboration grew, deciding which exchanges required formal review and how responsibility continued between projects.

The reviewing labour and allocation of credit remain open questions. Who would review an account that continued to change, and when would that responsibility pass to someone else? A contribution history could show who corrected an analysis or maintained an important comparison, but institutions would still have to judge the value of that work.

Open-source communities offer useful experience in organising such work. O'Mahony and Ferraro's (2007: 1082–1083, 1100) study of Debian shows how coordination and community building became valued leadership contributions, with formal authority subject to democratic checks. Franzoni and Sauermann (2014: 14–15) draw on open-source organisation to examine how scientific projects divide work and sustain leadership, while recognising the difficulty of fitting collective contributions into academic reward systems. CDMs could adapt these practices by giving responsibility and recognition to those who integrate contributions and keep the inquiry moving, alongside those who introduce new findings. Criteria for appointment and promotion in academia and universities would need to recognise such contributions if sustained participation in a CDM were to become a realistic part of an academic career. The challenge is to give consequential work appropriate weight without turning every recorded interaction into another metric to optimise.

Private science and crowdfunded research offer further possible avenues. Privately financed laboratories or independent researchers could bring evidence into a CDM, and a community could seek donations for a study emerging from its discussion. Scientific crowdfunding has already supported small projects, often involving junior researchers (Sauermann et al. 2019: 1). These routes raise questions about which materials sponsors would share and how financial backing would affect the choice of questions. Agreements about access and scientific decisions would determine how privately supported work could contribute to the shared account.

After unsuccessful grant applications, Wilson and colleagues used Experiment.com to fund a reference genome for the Gila monster lizard. Wilson (2019: 1, 3) describes how the campaign also built lasting connections with supporters and enabled further work on the animal's sex chromosomes. A CDM could give such a community a continuing place to discuss the research and develop questions for another funding round, connecting donations to an inquiry that outlasted the campaign.

Answers to these questions would depend on disciplinary cultures and regional arrangements for research. The Leiden Manifesto's attention to differences between fields remains useful here (Hicks et al. 2015: 429–431): a contribution that matters in one community may take a different form in another. Systems of appointment and promotion would also shape what researchers could afford to contribute. These differences leave room for several forms of CDM to develop, with communities negotiating how work undertaken in one setting could receive recognition elsewhere.

## 8 Conclusion

OpenAI's mathematics release and its subsequent academic collaborations show why expanding scientific production also calls for collective work to understand the results. Science has adapted to growth before, and the institutions that distributed its results also helped distribute authority. AI is likely to accelerate another such change. If scientific communities leave these choices to service providers, they may gain powerful tools while losing influence over how their work is organised.

Collaborative Debating Models offer one possible response: gathering places where co-intelligence becomes collective inquiry and participants develop an account of the work together. Connecting that inquiry to shared data and experimental services could also widen opportunities to produce new evidence. The four ownership routes give this possibility different institutional futures. Publishing houses could extend their journals into CDMs; scholarly associations and academic institutions could anchor them in established communities. Funding agencies could sustain them as public research infrastructure, while direct AI-company ownership would tie their future more closely to a provider's priorities. These choices would affect who could turn their questions into research and whose contributions entered the shared account. The contribution of the thought experiment is to show how greater productive capacity could be connected to continuing collective understanding through institutions that researchers help shape.

## Statements and Declarations

### Funding

The author received no specific funding for this work.

### Competing interests

The author declares no competing interests.

### Use of artificial intelligence

ChatGPT was used to assist literature searches and source comparison, as well as drafting and revision. The author is responsible for the final manuscript, including its arguments and interpretation of sources.

### Data availability

No new datasets were generated or analysed for this conceptual article.